%% file: superwind_ulx_v6_sub.tex
\documentclass[twocolumn,tighten,times]{aastex63}

\usepackage[T1]{fontenc}

\newcommand\ergs{erg~s$^{-1}$}

\newcommand{\xmm}{\textit{XMM-Newton}}
\newcommand{\nustar}{\textit{NuSTAR}}

\shorttitle{}
\shortauthors{}

\begin{document}

\title{Linking soft excess in ultraluminous X-ray sources with optically thick wind driven by supercritical accretion}

\author{Yanli Qiu}
\affiliation{Department of Astronomy, Tsinghua University, Beijing 100084, China}

\correspondingauthor{Hua Feng}
\email{hfeng@tsinghua.edu.cn}

\author[0000-0001-7584-6236]{Hua Feng}
\affiliation{Department of Astronomy, Tsinghua University, Beijing 100084, China}
\affiliation{Department of Engineering Physics and Center for Astrophysics, Tsinghua University, Beijing 100084, China}

\begin{abstract}
Supercritical accretion onto compact objects may drive massive winds that are nearly spherical, optically thick, and Eddington limited.  Blackbody emission from the photosphere is the direct observational signature of the wind. Here we investigate whether or not it can explain the soft emission component seen in the energy spectra of ultraluminous X-ray sources (ULXs). Based on high-quality \xmm\ spectra of 15 ULXs, we find that the soft component can be modeled as blackbody emission with a nearly constant luminosity, and the 5 known pulsating ULXs (PULXs) in the sample display a blackbody luminosity among the lowest. These are consistent with the scenario that the soft emission originates from the photosphere of the optically thick wind. However, the derived blackbody luminosity for PULXs is significantly above the Eddington limit for neutron stars. A possible explanation is that a considerable fraction of the optically thick wind roots in the inner accretion flow, where the radiative flux could exceed the Eddington limit due to a reduced scattering cross-section or enhanced radiation transfer with magnetic buoyancy.  Based on a wind model, the inferred mass accretion rate in these standard ULXs overlaps but is on average lower than that in luminous and very soft X-ray sources, which are also candidates with supercritical accretion. Alternatively, it cannot be ruled out that the soft emission component is a result of the hard component, e.g., via down-scattering in a cool medium, as a weak correlation may exist between them. 
\end{abstract}


\section{Introduction}

Ultraluminous X-ray sources (ULXs) are non-nuclear point-like X-ray sources with luminosities above the Eddington limit of typical stellar mass black holes \citep{Kaaret2017}.  Mounting evidence suggests that some or even the majority of them are powered by supercritical accretion onto compact objects, with the smoking gun being the identification of neutron stars in ULXs \citep{Bachetti2014,Carpano2018,Fuerst2016,Israel2017,Israel2017a,Tsygankov2017}.  Blue-shifted absorption lines are seen in ULXs \citep{Walton2016,Pinto2016,Pinto2017,Kosec2018}, suggesting the presence of ultrafast outflows (UFOs). Shock-ionized bubbles surrounding some ULXs are found to be produced by winds with an ultraluminous mechanical power interacting with the interstellar medium \citep{Pakull2003,Abolmasov2007,Belfiore2020}.  These winds/outflows are thought to be signatures associated with supercritical accretion.

The physics in the context of supercritical accretion is still poorly understood.  Based on numerical simulations in recent years,  a general picture is depicted that supercritical accretion will lead to a turbulent inflow in conjunction with strong winds \citep{Ohsuga2011,Jiang2014,Sadowski2016,Kitaki2018}. The central wind is optically thin, confined in a funnel like geometry with a velocity of $\sim$0.1--0.4~$c$, reminiscent of the UFO mentioned above.  The massive wind at large radii is speculated to be \textit{nearly spherical, optically thick, and Eddington-limited}~\citep{Meier1982_PaperII}, regarded as one of the unique signatures of supercritical accretion. Analytical analysis \citep{Shakura1973,Lipunova1999,King2003,Poutanen2007,Shen2016} predicts that, due to the presence of strong radiation pressure, the typical radius to lift the optically thick wind is roughly $\dot{m} R_{\rm isco}$, where $\dot{m}$ is the dimensionless mass accretion rate and $R_{\rm isco}$ is the radius of the innermost stable circular orbit. Unfortunately, none of the numerical works to date has a simulation box that is large enough to enable investigations of the entire optically thick wind. 

Observationally, the optically thick wind manifests itself simply as blackbody emission from its photosphere. Different from standard ULXs, supersoft ULXs display an energy spectrum dominated by a cool blackbody component, arguably due to emission from the photosphere of the optically thick wind driven by supercritical accretion \citep{Feng2016,Soria2016,Urquhart2016}.  They differ from standard ULXs and appear to be supersoft likely due to a large viewing angle such that the central hard X-rays are beamed away from the line of sight. Because the blackbody luminosity is expected at the Eddington limit,  \citet{Zhou2019} realized that similar accreting objects should appear very soft but not necessarily ultraluminous, as the Eddington limit of neutron stars is below the luminosity threshold of ULXs.  They thus assembled a catalog of luminous and very soft sources in nearby galaxies as candidates with supercritical accretion, and ruled out that they were nuclear burning white dwarfs or supernova remnants. Interestingly,  a possible bimodal distribution of the blackbody luminosity is revealed, with two peaks respectively around the Eddington limit of neutron stars and black holes \citep{Zhou2019}. Using a radiative hydrodynamic (RHD) wind model \citep{Meier1982_PaperII,Meier1982_PaperIII,Meier2012,Zhou2019}, the mass accretion rate is inferred to be hundreds of the critical rate.  Such a signature, blackbody emission around the Eddington limit, is also detected in an accreting pulsar Swift~J0243.6$+$6124 only when it is in the super-Eddington phase during an outburst \citep{Tao2019}. Furthermore, \citet{Yao2019} considered radiation transfer in the wind and proposed a wind-disk irradiation model, and found that it can well explain the X-ray, UV, and optical emission from a soft ULX in NGC~247.  

The standard ULXs, especially those containing a neutron star, are good laboratories for the study of supercritical accretion. Thus, it is of great interest to test the above wind model with them. The soft excess detected in the energy spectrum of standard ULXs has been proposed as emission due to the same physical origin \citep{Middleton2015, Weng2018}, but alternatives exist \citep{Barnard2010,Miller2013}.  Here, as an extension of \citet{Zhou2019}, we investigate the soft component in the energy spectrum of standard ULXs, to see whether it can be interpreted as emission from the photosphere of the optically thick wind driven by supercritical accretion.  The sample and observations are described in \S~\ref{sec:sample}, the spectral analysis and results are presented in \S~\ref{sec:results}, and the physical origin of the soft excess is discussed in \S~\ref{sec:discussion}.

\section{Sample and observations}
\label{sec:sample}

This work relies on ULXs with a soft excess in their energy spectrum, and requires high-quality data measured with the PN instrument on \xmm. Based on the \xmm\ ULX catalog \citep{Walton2011}, we select 24 ULXs with observations in which the PN instrument detects at least 5000 photons in the energy range of 0.3--10~keV from the 4XMM-DR9 catalog \citep{Webb2020}. This selection criterion is not strict, just in order to include all of the known PULXs and as many ULXs as possible in the first place. We discard 12 of them because the soft excess cannot be detected at a significance of $3\sigma$ or above in their energy spectra (see the spectral analysis in \S~\ref{sec:results}). 

As mentioned above, PULXs are of essential importance in this work,  because the mass of the compact object and consequently the Eddington limit is well constrained. Therefore, in addition to M51 ULX7 and ULX8, we add to our sample another 3 PULXs, {\it i.e.}, NGC 300 ULX1, NGC 1313 X-2, and NGC 7793 P13, which are not included in the ULX catalog; observations of these sources all satisfy the selection criteria.  We do not include M82 X-2 because it cannot be spatially resolved with \xmm.  In sum, a sample of 15 ULXs including 5 PULXs is constructed, with basic properties listed in Table \ref{tab:sample}. A total number of 49 PN observations are used in the analysis.  

The \texttt{sas} package version 16.1.0 with updated calibration files is adopted for data reduction.  The tasks \textit{epproc} are used to create calibrated and clean event lists.  Time intervals where the background flux is higher than 1.2 times the mean level are removed with the task \textit{espfilt} to exclude contaminations from soft protons. The source energy spectra are extracted from events with ${\rm FLAG} = 0$ and ${\rm PATTERN} \leq 4$, in a circular region centered on the source.  The source extraction radius is 40\arcsec\ in most cases but smaller if there is source confusion (listed in Table~\ref{tab:sample}). Background spectra are extracted from nearby source-free regions on the same chip; if there are multiple observations for the same target, background is obtained from the same regions.  The response files are created with tasks \textit{rmfgen} and \textit{arfgen}. The spectra are grouped to have at least 50 counts in each spectral bin. 

\section{Spectral analysis and results}
\label{sec:results}

A two-component model, blackbody plus Comptonization, is adopted to fit the spectrum using XSPEC version 12.11. The blackbody component is to account for the soft excess, \textit{i.e.}, emission presumably from the photosphere of the optically thick wind.  For the hard X-ray emission, the thermal Comptonization model \citep[\textsc{nthcomp};][]{Zdziarski1996, Zycki1999} is adopted following previous studies \citep[\textit{e.g.},][]{Middleton2015}.  We do not assume that the Comptonization seed photon temperature equals the blackbody temperature, as the Comptonization may take place in inner regions \citep{Jiang2014,Narayan2017} while the wind photosphere has a large radius \citep{Zhou2019}.  The whole spectrum is subject to interstellar absorption, modeled using two \textsc{tbabs} components, one fixed to account for the Milky Way absorption \citep{Kalberla2005} and the other for extragalactic and intrinsic absorption.

Some ULXs, like NGC 1313 X-1, NGC 5408 X-1 and NGC 55 ULX1, display blended atomic features in high-quality CCD spectra \citep{Middleton2014, Pinto2016, Walton2016}. Whether or not taking them into account in the fitting, we get consistent parameters for the continua within errors \citep[also see][]{Pinto2020}. Therefore, these minor features are not modeled in this work.

A hard X-ray convex at energies above a few keV is common in the energy spectra of ULXs. This feature is out of the interest of this study, but how the hard component is modeled may affect the determination of the soft component.  Not every target in our sample has available \nustar\ observations that allow us to constrain the hard curvature accurately.  We compare the fitting result with \nustar\ plus \xmm\ data versus that with \xmm\ data only for NGC 1313 X-1, where high-quality data with both telescopes are available. We find that the blackbody temperature and luminosity from the two fits differ by about 1$\sigma$, suggesting that \xmm\ data only should serve the purpose of this study.  

Another possible systematic error in the determination of the blackbody component may arise from the choice of the hard component. Besides \textsc{nthcomp}, we test with another Comptonization model (\textsc{comptt}) and the cutoff power-law model (\textsc{cutoffpl}). They offer comparable goodness of fit. No matter which one is used, consistent blackbody temperature and luminosity within errors are obtained, with difference mostly within 1$\sigma$. Thus, we conclude that the blackbody component can be robustly determined, and the systematic error is not larger than the statistical error. In this work, the \textsc{nthcomp} model is used for the hard emission.

The fitting results for each observation are listed in Table~\ref{tab:fit}. Besides the spectral parameters for the two components, the 0.3--10 keV luminosity of the Comptonization component ($L_{\rm Comp}$) and the whole model ($L_{\rm X}$), $\chi^2$, and the degree of freedom of the fit are also included.  We note that the blackbody luminosity is quoted as the bolometric luminosity, \textit{i.e.}, $L_{\rm bb} \equiv 4 \pi \sigma R_{\rm bb}^2 T_{\rm bb}^4$, where $R_{\rm bb}$ is the radius and $T_{\rm bb}$ is the effective temperature of the blackbody component, and $\sigma$ is Stefan-Boltzmann constant.

\begin{figure}[t]
\centering
\includegraphics[width=0.8\columnwidth]{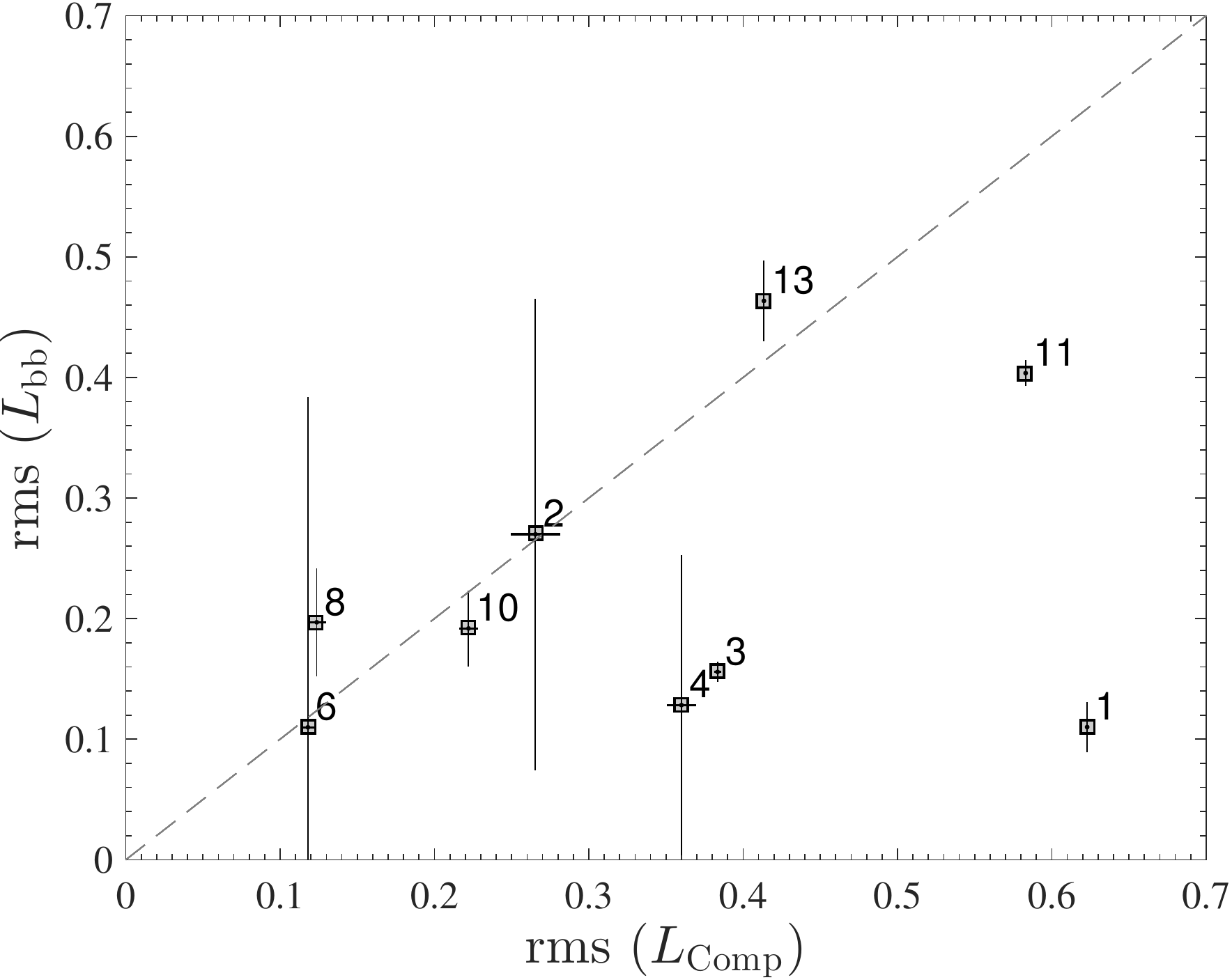}
\caption{Inter-observational fractional rms of the blackbody luminosity versus that of the Comptonization luminosity with 1$\sigma$ errors, for sources with at least 3 observations in our sample. The dashed line marks the 1:1 relation. The label indicates the source index listed in Table~\ref{tab:sample}. }
\label{fig:rms}
\end{figure}

First, we want to check whether or not the blackbody component displays a constant luminosity as expected.  For sources with at least 3 observations, the fractional root-mean-square (rms) of $L_{\rm bb}$ versus that of $L_{\rm Comp}$ with 1$\sigma$ errors is calculated following \citet{Vaughan2003} and shown in Figure~\ref{fig:rms}. As one can see, the Comptonization component is more variable than the blackbody component.  For most sources, the derived blackbody temperature and luminosity are found to have little variation (see Table~\ref{tab:fit}). Despite the measurement errors, the maximum to minimum luminosity ratio for the blackbody component is around 4 for Holmberg IX X-1 and less than 3 for others.  Holmberg IX X-1 is also the one with the highest rms variability ($0.46 \pm 0.03$). We then try to fit all of its observations jointly and impose only one additional condition that the source has a constant $L_{\rm bb}$ over different observations. The joint fit results in $\chi^2 = 602.0$ with 566 degrees of freedom, with a null hypothesis probability of 0.11, indicative of an acceptable fit (in general, one considers $>0.05$ being acceptable). Inspection of the residuals finds no significant change between independent fits and a joint fit.  The conclusion remains for other sources.  For NGC 1313 X-1, deep, multi-epoch broadband spectroscopy also reveals a constant blackbody luminosity \citep{Walton2020}. 

\begin{figure}[t]
\centering
\includegraphics[width=0.8\columnwidth]{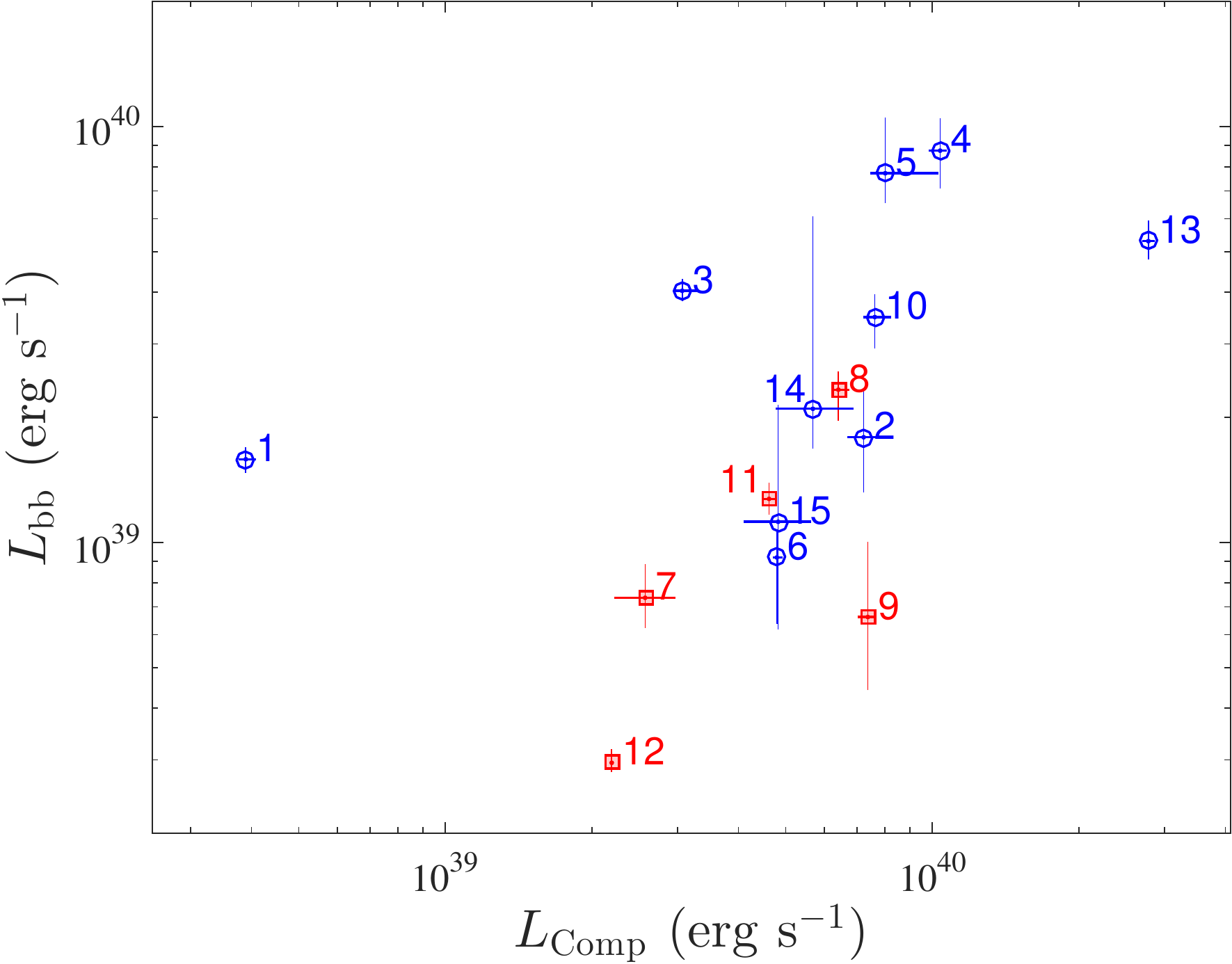}
\caption{Blackbody luminosity versus 0.3--10 keV Comptonization luminosity for sources in our sample.  The label indicates the source index listed in Table~\ref{tab:sample}. The median luminosity is adopted for sources with multiple observations. The error bar indicates the typical 1$\sigma$ uncertainty with a single observation. The red squares mark the known PULXs. }
\label{fig:lbb_lcomp_median}
\end{figure}

\begin{figure}[t]
\centering
\includegraphics[width=0.8\columnwidth]{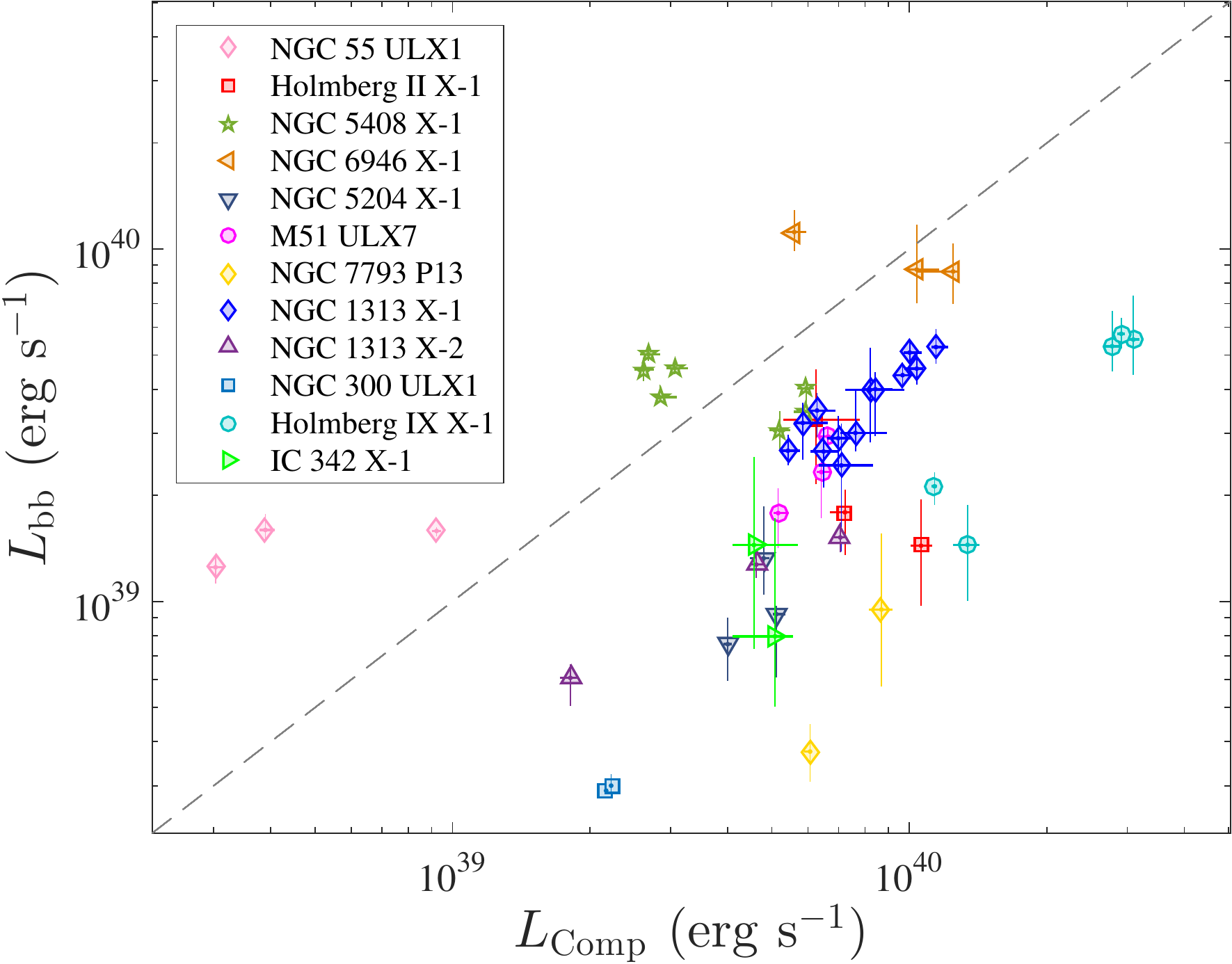}
\caption{Blackbody luminosity versus Comptonization luminosity derived from individual observations for sources with multiple observations in our sample with 1$\sigma$ errors. }
\label{fig:lbb_lcomp_all}
\end{figure}

The blackbody luminosity versus the 0.3--10 keV Comptonization luminosity for each source is displayed in Figure~\ref{fig:lbb_lcomp_median}.  For sources with multiple observations, the median luminosity with the typical measurement error is plotted, as one would see with a single observation.  It is obvious that the blackbody luminosity of PULXs is among the lowest.  However, we note that they are significantly higher than the Eddington limit of neutron stars. The same plot for individual observations are shown in Figure~\ref{fig:lbb_lcomp_all} for sources with multiple observations, to examine if there is any correlations between the two components. 

\begin{figure}[t]
\centering
\includegraphics[width=0.8\columnwidth]{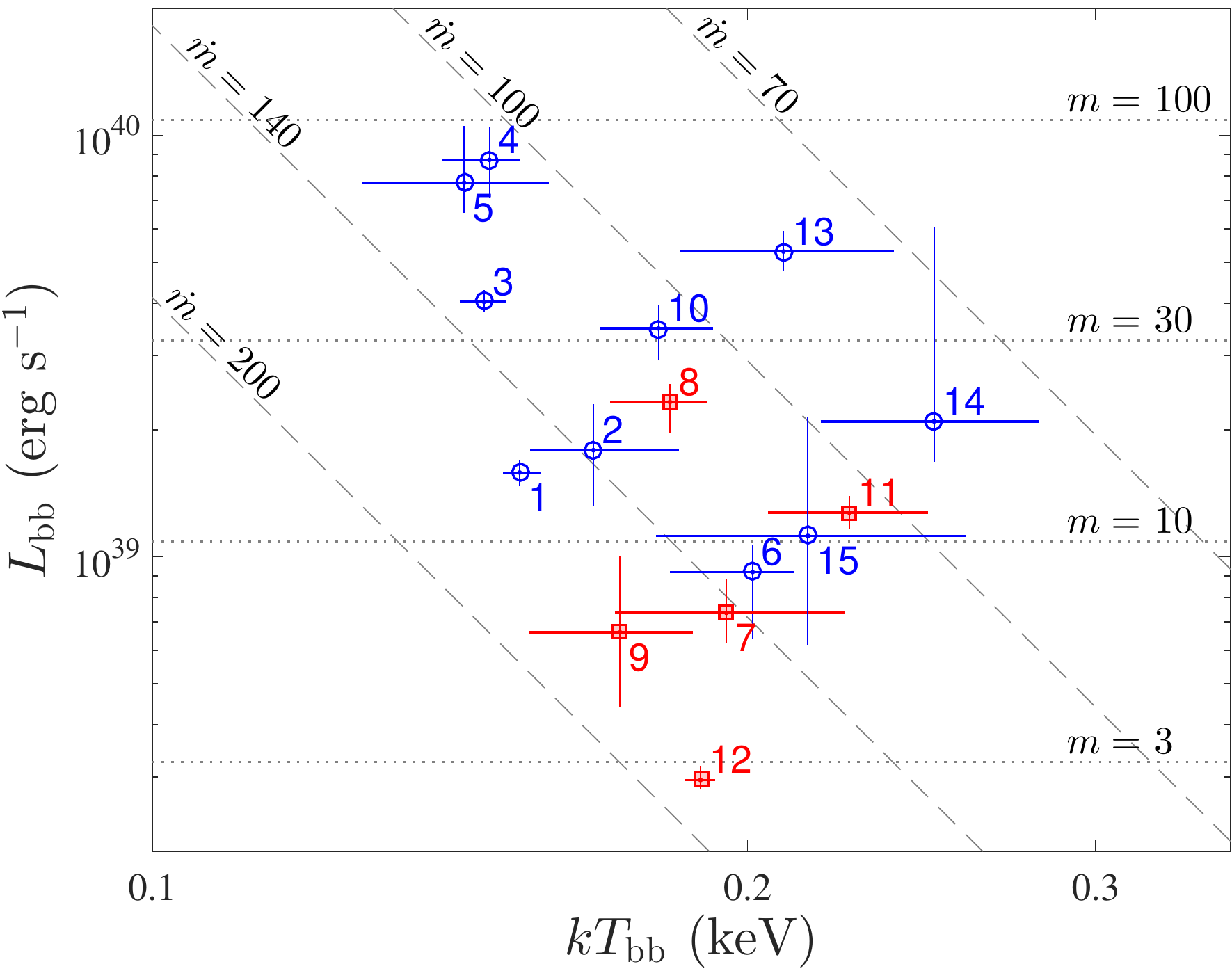}
\caption{Blackbody luminosity versus temperature for sources in our sample.  The median value is adopted for sources with multiple observations, and the error bar indicates the typical 1$\sigma$ uncertainty expected from a single observation.  The dotted lines mark the compact object mass and the dashed lines mark the mass accretion rate given a 1D RHD wind model. The blackbody luminosity is a function of the compact object mass, while the temperature decreases with increasing accretion rate. The label indicates the source index listed in Table~\ref{tab:sample}. The red squares mark the known PULXs. }
\label{fig:lbb_tbb}
\end{figure}

The observed properties of the wind photosphere driven by supercritical accretion can be predicted by a 1D RHD wind model \citep{Meier1982_PaperII}. The optically thick wind is assumed to launch at the advective radius of a slim disk \citep{Meier2012,Zhou2019},  and then develop through adiabatic acceleration and free expansion, respectively.  In the model, as long as the mass accretion rate is much higher than the supercritical rate ($\dot{m} \gg 1$), the blackbody luminosity is only determined by the mass of the compact object ($L_{\rm bb} = \frac{3}{4}L_{\rm Edd}$, given the above boundary conditions), while the blackbody temperature decreases with increasing accretion rate. In Figure~\ref{fig:lbb_tbb}, the median blackbody luminosity and temperature for each source are plotted against the model predictions. 

\section{Discussion}
\label{sec:discussion}

After a successful application of the optically thick wind model in luminous and very soft sources \citep{Zhou2019}, as well as in the interpretation of the multiwavelength spectrum of a supersoft ULX \citep{Yao2019}, here we investigate the physical origin of the soft excess in standard ULXs, to see if it can be self-consistently explained as due to thermal emission from the photosphere of the optically thick wind driven by supercritical accretion. One of the technical difficulties is that the soft excess in standard ULXs usually is not the dominant component. In most cases, the Comptonization component occupies the majority of the flux, and if not properly modeled, may in turn influence the characterization of the soft component. To this end, different Comptonization models are tested and we argue that the systematics on the derived blackbody luminosity and temperature are not larger than their statistical errors and can be ignored. Also, in this scenario, the emission from the disk outside the wind photosphere has a luminosity 2--3 orders of magnitude lower than that from the wind photosphere, and is thus negligible.

The lack of variability in the soft thermal component is consistent with previous studies \citep{Middleton2015,Weng2018,Walton2020}. It is in line with the assumption that the thermal emission arises from the wind photosphere. In the scenario of the optically thick wind,  the blackbody luminosity is a constant and levels at the Eddington limit. This may have little dependence on detailed models, and can be understood as follows. The wind is driven by the radiation pressure, so that the typical radius where the disk puffs up and the wind takes off is the radius where the radiation approaches the Eddington limit. Under the assumption of local thermodynamic equilibrium, the outward radiative flux conserves along with the development of the wind, and escapes as blackbody emission with a luminosity equal to that at the inner boundary, \textit{i.e.}, around the Eddington limit. This implies that the wind thermal emission could potentially be used as an indicator of the compact object mass in the case of supercritical accretion. 

The PULXs in our sample display blackbody luminosities among the lowest, although their Comptonization luminosities are typical in the sample. However, the measured blackbody luminosities of the PULXs, except NGC 300 ULX1 (source 12), is systematically higher than the Eddington limit of neutron stars assuming spherical accretion, \textit{i.e.}, $(1-3) \times 10^{38}$~\ergs.  Due to the presence of strong magnetic fields, the electron scattering opacity could be largely depressed, leading to a higher Eddington limit \citep{Mushtukov2015,Mushtukov2016}. In the model, the wind is assumed to launch from a single, typical radius, while in reality the wind base should cover a wide range of radii. In inner regions, where magnetic buoyancy may play an important role in the transportation of radiation \citep{Socrates2006,Blaes2011,Jiang2014}, the wind luminosity could exceed the Eddington limit significantly due to such an effect. 

However, other candidates with supercritical accretion do not seem to be special in their blackbody luminosities. A tentative peak (or enhancement) around $10^{38}$~\ergs\ is seen in the distribution of blackbody luminosities of very/super soft X-rays sources \citep{Zhou2019}. A blackbody component with a luminosity of $(1-2) \times 10^{38}$~\ergs\ only appears during the super-Eddington phase of the Galactic accreting pulsar Swift~J0243.6$+$6124 \citep{Tao2019}.  The mass accretion rate of sources in our sample is inferred to be in the range of $\dot{m} = 100 - 200$, which is broadly consistent with the rate inferred from very/super soft sources ($\dot{m} = 100 - 500$) based on the same model, but is on average lower.  Supersoft ULXs are perhaps systems viewed at high inclinations such that the hard X-rays from the central are largely steered away from our line of sight \citep{Soria2016,Urquhart2016}.  It has also been argued that supersoft ULXs may have higher accretion rates than standard ULXs \citep{Feng2016,Zhou2019}, to account for the facts that their blackbody temperatures are systematically lower and state transitions have been seen in NGC 247 X-1 between the supersoft and standard ultraluminous state.  Therefore, their difference in mass accretion rate may explain why the wind photosphere in standard ULXs is systematically more luminous.  Another possibility could be that the strength of the neutron star magnetic fields is different between the two populations \citep{Erkut2020}. 

On the other hand, we note that  sources in our sample may show a possible correlation  between $L_{\rm bb}$ and $L_{\rm Comp}$.  The Spearman's correlation coefficient between the median $L_{\rm bb}$ and $L_{\rm Comp}$ is 0.64 with a $p$-value of 0.01, indicative of a weak correlation, if any.  If the correlation really exists, it means that the blackbody emission may respond to the hard component, \textit{e.g.}, due to down-scattering of hard X-rays in a cooler medium \citep{Kawashima2012}. Detection of soft lags suggests that a fraction of the soft X-rays may be due to such a process \citep{Pinto2017,Kara2020}. For individual objects, such a correlation is not shown in sources like NGC 55 ULX1 and NGC 5408 X-1, but could exist in some others (see Figure~\ref{fig:lbb_lcomp_all}).  Thus, we conclude that this interpretation is not well supported by observations but could still be possible. 

To conclude, the optically thick wind scenario can explain most of the observations with the soft excess in standard ULXs, but there are obvious caveats, in particular the deviation between the blackbody luminosity and the typical Eddington limit. This deviation also prevents us from using the blackbody luminosity to estimate the compact object mass in standard ULXs. To fully understand the properties of the optically thick wind, future numerical simulations with a box large enough (stable up to $\dot{m}R_{\rm isco}$) are needed.  

\acknowledgments We thank the anonymous referee for useful comments. HF acknowledges funding support from the National Key R\&D Project (grants Nos.\ 2018YFA0404502 \& 2016YFA040080X), and the National Natural Science Foundation of China under the grant Nos.\ 11633003 \& 11821303.  

\begin{deluxetable*}{llcccccccl}
\tabletypesize{\footnotesize}
\tablecaption{The ULX sample in this work. }
\label{tab:sample}
\tablehead{
\colhead{No.} & \colhead{Name} & \colhead{R.A.} & \colhead{decl.} &  \colhead{$N_{\rm H, Gal}$}  & \colhead{$d$}  & \colhead{ref.}   & \colhead{$R_{\rm src}$} &  \colhead{$N_{\rm obs}$}   &  \colhead{$L_{\rm bb}$}   \\
\colhead{} & \colhead{} & \colhead{(J2000)} & \colhead{(J2000)} &  \colhead{($10^{21}$ cm$^{-2}$)}  & \colhead{(Mpc)}  & \colhead{}   & \colhead{(\arcsec)}   & \colhead{}   & \colhead{($10^{38}$ $\rm erg~s^{-1}$)}   \\
\colhead{(1)} & \colhead{(2)} & \colhead{(3)} & \colhead{(4)} &  \colhead{(5)}  & \colhead{(6)}  & \colhead{(7)}   & \colhead{(8)}   & \colhead{(9)}   & \colhead{(10)} 
} 
\startdata
1 & NGC 55 ULX1 & 00:15:28.9 & $-$39:13:18 & 0.94 & 1.8 & 1 & 40 & 3 & $ 15.8 ^{+ 1.1 } _{- 1.2 } $ \\
2 & Holmberg II X-1 & 08:19:28.9 & $+$70:42:19 & 0.57 & 3.3 & 2 & 40 & 3 & $ 17.9 ^{+ 5.1 } _{- 4.7 } $ \\
3 & NGC 5408 X-1 & 14:03:19.6 & $-$41:22:58 & 0.52 & 4.8 & 3 & 40 & 7 & $ 40.3 ^{+ 2.7 } _{- 2.2 } $ \\
4 & NGC 6946 X-1 & 20:35:00.7 & $+$60:11:30 & 2.20 & 7.7 & 4 & 40 & 3 & $ 87.5 ^{+ 17.0 } _{- 17.4 } $ \\
5 & NGC 4559 ULX1 & 12:35:51.7 & $+$27:56:04 & 0.14 & 9.7 & 5 & 40 & 1 & $ 77.2 ^{+ 28.0 } _{- 11.7 } $ \\
6 & NGC 5204 X-1 & 13:29:38.6 & $+$58:25:06 & 0.26 & 4.9 & 2 & 40 & 3 & $ 9.2 ^{+ 1.4 } _{- 2.8 } $ \\
7 & M51 ULX8$^\ast$ & 13:30:07.6 & $+$47:11:06 & 0.33 & 8.6 & 6 & 40 & 1 & $ 7.4 ^{+ 1.5 } _{- 1.1 } $ \\
8 & M51 ULX7$^\ast$ & 13:30:01.0 & $+$47:13:43 & 0.33 & 8.6 & 6 & 40 & 3 & $ 23.3 ^{+ 2.5 } _{- 3.7 } $ \\
9 & NGC 7793 P13$^\ast$ & 23:57:50.9 & $-$32:37:27 & 0.33 & 3.6 & 2 & 30 & 2 & $ 6.6 ^{+ 3.4 } _{- 2.2 } $ \\
10 & NGC1313 X-1 & 03:18:20.0 & $-$66:29:11 & 4.00 & 4.3 & 2 & 40 & 13 & $ 34.8 ^{+ 4.7 } _{- 5.6 } $ \\
11 & NGC1313 X-2$^\ast$ & 03:18:22.1 & $-$66:36:03 & 0.70 & 4.3 & 2 & 40 & 3 & $ 12.7 ^{+ 1.2 } _{- 1.1 } $ \\
12 & NGC 300 ULX1$^\ast$ & 00:55:04.8 & $-$37:41:43 & 0.93 & 1.9 & 7 & 30 & 2 & $ 3.0 \pm 0.2 $ \\
13 & Holmberg IX X-1 & 09:57:53.2 & $+$69:03:48 & 0.81 & 3.8 & 2 & 40 & 5 & $ 53.0 ^{+ 6.4 } _{- 5.2 } $ \\
14 & NGC 2403 X-1 & 07:36:25.5 & $+$65:35:39 & 1.13 & 4.2 & 5 & 40 & 1 & $ 21.0 ^{+ 39.8 } _{- 4.1 } $ \\
15 & IC 342 X-1 & 03:45:55.6 & $+$68:04:55 & 3.62 & 3.4 & 2 & 32 & 2 & $ 11.2 ^{+ 10.2 } _{- 5.0 } $ \\
\enddata
\tablecomments{
Column~1: Object ID.
Column~2: Source name; the $^\ast$ symbol denotes a PULX.
Column~3: Right ascension quoted in 4XMM-DR9. 
Column~4: Declination quoted in 4XMM-DR9. 
Column~5: Absorption column density along the line of sight to the source \citep{Kalberla2005}.
Column~6: Distance to the host galaxy.
Column~7: Reference for the distance. 
Ref.~1: \citet{Karachentsev2003}; 
Ref.~2: \citet{Tully2016}; 
Ref.~3: \citet{Karachentsev2002a}; 
Ref.~4: \citet{Anand2018}; 
Ref.~5: \citet{Ho1997};  
Ref.~6: \citet{McQuinn2016}; 
Ref.~7: \citet{Gieren2005}.
Column~8: Source extraction radius. 
Column~9: Number of observations used in this work. 
Column~10: The median blackbody luminosity.
}
\end{deluxetable*}

\clearpage

\input{table2.tex}


\clearpage


\end{document}

%% file: table2.tex
   \startlongtable
\begin{deluxetable*}{lccccccccc}
\tablewidth{\textwidth}
\tablecolumns{10}
\tabletypesize{\scriptsize}
\tablecaption{Spectral fitting for each observation. \label{tab:fit}}
\tablehead{
\colhead{ObsID} & \colhead{$N_{\rm H}$ } & \colhead{$kT_{\rm bb}$}  & \colhead{$L_{\rm bb}$}   & \colhead{$\Gamma$}  & \colhead{$kT_{\rm e}$} & \colhead{$kT_0$} & \colhead{$L_{\rm Comp}$} & \colhead{$L_{\rm X}$}  & \colhead{$\chi^2/$dof}\\
\colhead{} & \colhead{($10^{21}$ cm$^{-2})$} & \colhead{(keV)} & \colhead{$(10^{38}$ \ergs)} & \colhead{} & \colhead{(keV)} & \colhead{(keV)} & \colhead{($10^{39}$ \ergs)} & \colhead{($10^{39}$ \ergs)} & \colhead{} \\
\colhead{(1)} & \colhead{(2)} & \colhead{(3)} & \colhead{(4)} & \colhead{(5)} & \colhead{(6)} & \colhead{(7)} & \colhead{(8)} & \colhead{(9)} & \colhead{(10)}
}
\startdata
\multicolumn{10}{c}{(1) NGC 55 ULX1}\\
\noalign{\smallskip}\hline\noalign{\smallskip}
   0028740201  &   $  1.22 ^{+ 0.15 } _{- 0.16 }  $     &   $  0.179 ^{+ 0.003 } _{- 0.002 }  $     &   $  15.8 ^{+ 0.4 } _{- 0.6 }  $      &   $  4.06 ^{+ 1.10 } _{- 0.42 }  $    &   $  1.7 ^{+ \infty } _{- 1.7 }  $     &   $  0.38 ^{+ 0.03 } _{- 0.07 }  $     &   $  0.92 ^{+ 0.02 } _{- 0.02 }  $    &   $  2.33 ^{+ 0.02 } _{- 0.02 }  $   &     81.1/77 
 \\    0655050101  &   $  1.87 ^{+ 0.19 } _{- 0.06 }  $     &   $  0.153 ^{+ 0.004 } _{- 0.005 }  $     &   $  16.0 ^{+ 1.7 } _{- 1.2 }  $      &   $  5.92 ^{+ 1.56 } _{- 0.98 }  $    &   $  4.9 ^{+ \infty } _{- 4.9 }  $     &   $  0.38 ^{+ 0.02 } _{- 0.02 }  $     &   $  0.39 ^{+ 0.02 } _{- 0.01 }  $    &   $  1.7 ^{+ 0.1 } _{- 0.1 }  $   &    150.3/81 
 \\    0824570101  &   $  2.00 ^{+ 0.15 } _{- 0.16 }  $     &   $  0.149 ^{+ 0.005 } _{- 0.003 }  $     &   $  12.5 ^{+ 1.1 } _{- 1.2 }  $      &   $  4.33 ^{+ 1.20 } _{- 0.33 }  $    &   $  2.0 ^{+ \infty } _{- 2.0 }  $     &   $  0.37 ^{+ 0.03 } _{- 0.02 }  $     &   $  0.30 ^{+ 0.01 } _{- 0.01 }  $    &   $  1.3 ^{+ 0.1 } _{- 0.1 }  $   &    137.2/80 
 \\ \cutinhead{(2) Holmberg II X-1} 
   0200470101  &   $  0.55 ^{+ 0.17 } _{- 0.17 }  $     &   $  0.184 ^{+ 0.012 } _{- 0.012 }  $     &   $  14.4 ^{+ 5.1 } _{- 4.7 }  $      &   $  2.34 ^{+ 0.05 } _{- 0.05 }  $    &   $  2.4 ^{+ 0.4 } _{- 0.3 }  $     &   $  0.09 ^{+ 0.02 } _{- 0.02 }  $     &   $  10.6 ^{+ 0.6 } _{- 0.5 }  $    &   $  13.2 ^{+ 0.7 } _{- 0.7 }  $   &   122.9/114 
 \\    0724810101  &   $  0.32 ^{+ 0.49 } _{- 0.32 }  $     &   $  0.150 ^{+ 0.018 } _{- 0.013 }  $     &   $  32.8 ^{+ 12.8 } _{- 11.3 }  $      &   $  1.89 ^{+ 0.20 } _{- 0.09 }  $    &   $  2.1 ^{+ 3.6 } _{- 0.4 }  $     &   $  0.19 ^{+ 0.19 } _{- 0.19 }  $     &   $  6.3 ^{+ 1.5 } _{- 1.0 }  $    &   $  9.0 ^{+ 1.5 } _{- 0.8 }  $   &     45.3/56 
 \\    0724810301  &   $  0.22 ^{+ 0.08 } _{- 0.22 }  $     &   $  0.167 ^{+ 0.019 } _{- 0.007 }  $     &   $  17.9 ^{+ 2.9 } _{- 4.4 }  $      &   $  1.88 ^{+ 0.07 } _{- 0.08 }  $    &   $  2.2 ^{+ 0.8 } _{- 0.2 }  $     &   $  0.01 ^{+ 0.09 } _{- 0.01 }  $     &   $  7.2 ^{+ 0.2 } _{- 0.5 }  $    &   $  8.8 ^{+ 0.5 } _{- 0.8 }  $   &     60.6/65 
 \\   \cutinhead{(3) NGC 5408 X-1} 
   0302900101  &   $  0.15 ^{+ 0.06 } _{- 0.06 }  $     &   $  0.142 ^{+ 0.003 } _{- 0.003 }  $     &   $  50.3 ^{+ 2.4 } _{- 2.2 }  $      &   $  2.80 ^{+ 0.18 } _{- 0.23 }  $    &   $  6.8 ^{+ \infty } _{- 4.4 }  $     &   $  0.30 ^{+ 0.03 } _{- 0.04 }  $     &   $  2.7 ^{+ 0.2 } _{- 0.1 }  $    &   $  6.7 ^{+ 0.2 } _{- 0.2 }  $   &    162.5/98 
 \\    0500750101  &   $  0.15 ^{+ 0.10 } _{- 0.10 }  $     &   $  0.147 ^{+ 0.005 } _{- 0.005 }  $     &   $  45.5 ^{+ 3.8 } _{- 3.2 }  $      &   $  2.93 ^{+ 0.14 } _{- 0.12 }  $    &          100     &   $  0.34 ^{+ 0.03 } _{- 0.03 }  $     &   $  2.6 ^{+ 0.1 } _{- 0.1 }  $    &   $  6.3 ^{+ 0.3 } _{- 0.3 }  $   &    114.8/75 
 \\    0653380201  &   $  0.44 ^{+ 0.03 } _{- 0.11 }  $     &   $  0.145 ^{+ 0.004 } _{- 0.003 }  $     &   $  40.3 ^{+ 2.6 } _{- 3.7 }  $      &   $  2.27 ^{+ 0.06 } _{- 0.05 }  $    &   $  1.7 ^{+ 0.2 } _{- 0.2 }  $     &   $  0.04 ^{+ 0.03 } _{- 0.04 }  $     &   $  6.0 ^{+ 0.2 } _{- 0.3 }  $    &   $  7.7 ^{+ 0.3 } _{- 0.3 }  $   &    164.1/96 
 \\    0653380301  &   $  0.39 ^{+ 0.05 } _{- 0.08 }  $     &   $  0.147 ^{+ 0.004 } _{- 0.003 }  $     &   $  34.6 ^{+ 1.6 } _{- 1.8 }  $      &   $  2.32 ^{+ 0.05 } _{- 0.05 }  $    &   $  1.9 ^{+ 0.2 } _{- 0.2 }  $     &   $  0.02 ^{+ 0.04 } _{- 0.02 }  $     &   $  5.9 ^{+ 0.5 } _{- 0.3 }  $    &   $  8.8 ^{+ 0.3 } _{- 0.3 }  $   &   186.9/108 
 \\    0653380401  &   $  0.05 ^{+ 0.08 } _{- 0.05 }  $     &   $  0.149 ^{+ 0.003 } _{- 0.004 }  $     &   $  45.9 ^{+ 2.7 } _{- 1.7 }  $      &   $  2.84 ^{+ 0.08 } _{- 0.07 }  $    &          100     &   $  0.31 ^{+ 0.02 } _{- 0.02 }  $     &   $  3.1 ^{+ 0.2 } _{- 0.1 }  $    &   $  6.8 ^{+ 0.3 } _{- 0.2 }  $   &   159.2/100 
 \\    0653380501  &   $  0.30 ^{+ 0.13 } _{- 0.13 }  $     &   $  0.150 ^{+ 0.004 } _{- 0.005 }  $     &   $  30.6 ^{+ 4.1 } _{- 3.6 }  $      &   $  2.29 ^{+ 0.06 } _{- 0.06 }  $    &   $  2.2 ^{+ 0.6 } _{- 0.3 }  $     &   $  0.07 ^{+ 0.02 } _{- 0.07 }  $     &   $  5.2 ^{+ 0.3 } _{- 0.3 }  $    &   $  7.7 ^{+ 0.4 } _{- 0.4 }  $   &   118.8/103 
 \\    0723130301  &   $  0.00 ^{+ 0.12 } _{- 0.00 }  $     &   $  0.149 ^{+ 0.001 } _{- 0.007 }  $     &   $  38.0 ^{+ 3.2 } _{- 1.2 }  $      &   $  2.64 ^{+ 0.41 } _{- 0.25 }  $    &   $  2.8 ^{+ \infty } _{- 1.2 }  $     &   $  0.32 ^{+ 0.05 } _{- 0.06 }  $     &   $  2.8 ^{+ 0.3 } _{- 0.0 }  $    &   $  5.9 ^{+ 0.3 } _{- 0.0 }  $   &      115/70 
 \\   \cutinhead{(4) NGC 6946 X-1} 
   0401360101  &   $  0.68 ^{+ 0.34 } _{- 0.27 }  $     &   $  0.148 ^{+ 0.005 } _{- 0.011 }  $     &   $  87.5 ^{+ 29.6 } _{- 17.4 }  $      &   $  2.31 ^{+ 0.21 } _{- 0.13 }  $    &   $  2.7 ^{+ 8.4 } _{- 0.7 }  $     &   $  0.02 ^{+ 0.05 } _{- 0.02 }  $     &   $  10.4 ^{+ 1.2 } _{- 0.6 }  $    &   $  17.9 ^{+ 2.6 } _{- 2.0 }  $   &     73.7/66 
 \\    0691570101  &   $  0.60 ^{+ 0.23 } _{- 0.21 }  $     &   $  0.161 ^{+ 0.007 } _{- 0.008 }  $     &   $  111.7 ^{+ 17.0 } _{- 13.0 }  $      &   $  2.36 ^{+ 0.38 } _{- 0.18 }  $    &   $  2.7 ^{+ \infty } _{- 0.7 }  $     &   $  0.36 ^{+ 0.06 } _{- 0.07 }  $     &   $  5.6 ^{+ 0.3 } _{- 0.2 }  $    &   $  15.1 ^{+ 1.5 } _{- 1.2 }  $   &    138.4/96 
 \\    0794581201  &   $  0.68 ^{+ 0.03 } _{- 0.03 }  $     &   $  0.148 ^{+ 0.002 } _{- 0.002 }  $     &   $  86.4 ^{+ 17.3 } _{- 16.6 }  $      &   $  2.30 ^{+ 0.04 } _{- 0.03 }  $    &   $  2.4 ^{+ 1.0 } _{- 0.3 }  $     &   $  0.02 ^{+ 0.06 } _{- 0.02 }  $     &   $  12.5 ^{+ 0.3 } _{- 2.0 }  $    &   $  19.6 ^{+ 1.6 } _{- 2.4 }  $   &     75.8/69 
 \\  \cutinhead{(5) NGC 4559 ULX1} 
   0152170501  &   $  1.22 ^{+ 0.40 } _{- 0.35 }  $     &   $  0.144 ^{+ 0.015 } _{- 0.016 }  $     &   $  77.2 ^{+ 28.0 } _{- 11.7 }  $      &   $  2.07 ^{+ 0.33 } _{- 0.14 }  $    &   $  2.2 ^{+ \infty } _{- 0.5 }  $     &   $  0.28 ^{+ 0.09 } _{- 0.17 }  $     &   $  8.0 ^{+ 2.3 } _{- 0.5 }  $    &   $  14.2 ^{+ 2.5 } _{- 1.6 }  $   &       54/59 
 \\   \cutinhead{(6) NGC 5204 X-1} 
   0405690501  &   $  0.00 ^{+ 0.08 } _{- 0.00 }  $     &   $  0.216 ^{+ 0.010 } _{- 0.031 }  $     &   $  9.2 ^{+ 0.5 } _{- 3.1 }  $      &   $  1.97 ^{+ 0.07 } _{- 0.06 }  $    &          100     &   $  0.11 ^{+ 0.02 } _{- 0.02 }  $     &   $  5.1 ^{+ 0.2 } _{- 0.1 }  $    &   $  6.0 ^{+ 0.2 } _{- 0.0 }  $   &     78.7/72 
 \\    0693851401  &   $  0.53 ^{+ 0.29 } _{- 0.22 }  $     &   $  0.151 ^{+ 0.015 } _{- 0.018 }  $     &   $  13.3 ^{+ 5.3 } _{- 2.8 }  $      &   $  1.79 ^{+ 0.08 } _{- 0.08 }  $    &   $  2.0 ^{+ 0.5 } _{- 0.3 }  $     &   $  0.04 ^{+ 0.04 } _{- 0.04 }  $     &   $  4.8 ^{+ 0.1 } _{- 0.3 }  $    &   $  5.8 ^{+ 0.3 } _{- 0.3 }  $   &     71.3/57 
 \\    0741960101  &   $  0.00 ^{+ 0.17 } _{- 0.00 }  $     &   $  0.201 ^{+ 0.009 } _{- 0.018 }  $     &   $  7.6 ^{+ 1.4 } _{- 1.6 }  $      &   $  1.76 ^{+ 0.07 } _{- 0.06 }  $    &   $  2.1 ^{+ 0.5 } _{- 0.3 }  $     &   $  0.07 ^{+ 0.03 } _{- 0.07 }  $     &   $  4.0 ^{+ 0.1 } _{- 0.1 }  $    &   $  4.9 ^{+ 0.2 } _{- 0.2 }  $   &     64.9/65 
 \\       \cutinhead{(7) M51 ULX8} 
   0830191601  &   $  0.43 ^{+ 0.29 } _{- 0.43 }  $     &   $  0.195 ^{+ 0.029 } _{- 0.024 }  $     &   $  7.4 ^{+ 1.5 } _{- 1.1 }  $      &   $  1.83 ^{+ 0.16 } _{- 0.10 }  $    &   $  1.9 ^{+ 6.0 } _{- 0.5 }  $     &   $  0.01 ^{+ 0.13 } _{- 0.01 }  $     &   $  2.6 ^{+ 0.4 } _{- 0.3 }  $    &   $  2.9 ^{+ 0.2 } _{- 0.1 }  $   &     44.7/50 
 \\       \cutinhead{(8) M51 ULX7} 
   0824450901  &   $  0.00 ^{+ 0.21 } _{- 0.00 }  $     &   $  0.178 ^{+ 0.008 } _{- 0.010 }  $     &   $  17.8 ^{+ 3.2 } _{- 3.7 }  $      &   $  1.53 ^{+ 0.07 } _{- 0.02 }  $    &   $  2.0 ^{+ 0.3 } _{- 0.2 }  $     &   $  0.25 ^{+ 0.06 } _{- 0.25 }  $     &   $  5.2 ^{+ 0.3 } _{- 0.1 }  $    &   $  6.7 ^{+ 0.3 } _{- 0.1 }  $   &    108.6/93 
 \\    0830191501  &   $  0.26 ^{+ 0.19 } _{- 0.17 }  $     &   $  0.183 ^{+ 0.011 } _{- 0.012 }  $     &   $  29.4 ^{+ 1.9 } _{- 3.1 }  $      &   $  1.65 ^{+ 0.08 } _{- 0.06 }  $    &   $  2.9 ^{+ 1.7 } _{- 0.6 }  $     &   $  0.39 ^{+ 0.09 } _{- 0.12 }  $     &   $  6.6 ^{+ 0.4 } _{- 0.3 }  $    &   $  9.2 ^{+ 0.4 } _{- 0.4 }  $   &    110.6/92 
 \\    0830191601  &   $  0.00 ^{+ 0.18 } _{- 0.00 }  $     &   $  0.191 ^{+ 0.005 } _{- 0.016 }  $     &   $  23.3 ^{+ 2.5 } _{- 6.1 }  $      &   $  1.62 ^{+ 0.08 } _{- 0.09 }  $    &   $  2.1 ^{+ 0.8 } _{- 0.2 }  $     &   $  0.39 ^{+ 0.11 } _{- 0.21 }  $     &   $  6.4 ^{+ 0.4 } _{- 0.2 }  $    &   $  8.5 ^{+ 0.3 } _{- 0.2 }  $   &    102.1/92 
 \\   \cutinhead{(9) NGC 7793 P13} 
   0804670601  &   $  1.13 ^{+ 0.48 } _{- 0.43 }  $     &   $  0.148 ^{+ 0.020 } _{- 0.016 }  $     &   $  9.5 ^{+ 6.1 } _{- 3.7 }  $      &   $  1.43 ^{+ 0.02 } _{- 0.02 }  $    &   $  2.1 ^{+ 0.1 } _{- 0.1 }  $     &   $  0.10 ^{+ 0.26 } _{- 0.10 }  $     &   $  8.7 ^{+ 0.5 } _{- 0.5 }  $    &   $  10.6 ^{+ 0.7 } _{- 0.7 }  $   &   125.7/106 
 \\    0804670701  &   $  0.79 ^{+ 0.14 } _{- 0.19 }  $     &   $  0.197 ^{+ 0.011 } _{- 0.018 }  $     &   $  3.8 ^{+ 0.7 } _{- 0.7 }  $      &   $  1.42 ^{+ 0.02 } _{- 0.02 }  $    &   $  2.2 ^{+ 0.1 } _{- 0.1 }  $     &   $  0.01 ^{+ 0.08 } _{- 0.01 }  $     &   $  6.06 ^{+ 0.04 } _{- 0.17 }  $    &   $  6.4 ^{+ 0.1 } _{- 0.1 }  $   &   134.5/131 
 \\  \cutinhead{(10) NGC 1313 X-1} 
   0405090101  &   $  1.61 ^{+ 0.28 } _{- 0.22 }  $     &   $  0.167 ^{+ 0.012 } _{- 0.013 }  $     &   $  32.1 ^{+ 4.5 } _{- 6.9 }  $      &   $  1.72 ^{+ 0.04 } _{- 0.03 }  $    &   $  2.3 ^{+ 0.2 } _{- 0.2 }  $     &   $  0.29 ^{+ 0.08 } _{- 0.19 }  $     &   $  5.8 ^{+ 0.8 } _{- 0.3 }  $    &   $  8.6 ^{+ 0.5 } _{- 0.4 }  $   &   136.7/134 
 \\    0693850501  &   $  1.38 ^{+ 0.09 } _{- 0.18 }  $     &   $  0.184 ^{+ 0.012 } _{- 0.008 }  $     &   $  29.1 ^{+ 4.6 } _{- 5.9 }  $      &   $  1.77 ^{+ 0.03 } _{- 0.03 }  $    &   $  2.6 ^{+ 0.3 } _{- 0.2 }  $     &   $  0.24 ^{+ 0.09 } _{- 0.10 }  $     &   $  7.0 ^{+ 0.5 } _{- 0.4 }  $    &   $  9.5 ^{+ 0.4 } _{- 0.3 }  $   &   142.4/133 
 \\    0693851201  &   $  1.73 ^{+ 0.35 } _{- 0.26 }  $     &   $  0.173 ^{+ 0.008 } _{- 0.008 }  $     &   $  30.1 ^{+ 9.7 } _{- 3.4 }  $      &   $  1.80 ^{+ 0.02 } _{- 0.03 }  $    &   $  3.3 ^{+ 0.7 } _{- 0.4 }  $     &   $  0.15 ^{+ 0.16 } _{- 0.07 }  $     &   $  7.6 ^{+ 1.3 } _{- 0.6 }  $    &   $  10.6 ^{+ 0.5 } _{- 0.4 }  $   &   147.3/136 
 \\    0722650101  &   $  1.26 ^{+ 0.39 } _{- 0.52 }  $     &   $  0.195 ^{+ 0.035 } _{- 0.028 }  $     &   $  24.3 ^{+ 7.6 } _{- 10.5 }  $      &   $  1.75 ^{+ 0.11 } _{- 0.08 }  $    &          100     &   $  0.30 ^{+ 0.17 } _{- 0.19 }  $     &   $  7.1 ^{+ 1.2 } _{- 0.8 }  $    &   $  9.0 ^{+ 1.2 } _{- 0.5 }  $   &     47.6/63 
 \\    0742490101  &   $  1.37 ^{+ 0.23 } _{- 0.20 }  $     &   $  0.188 ^{+ 0.012 } _{- 0.013 }  $     &   $  26.8 ^{+ 2.9 } _{- 2.4 }  $      &   $  1.81 ^{+ 0.03 } _{- 0.05 }  $    &   $  8.1 ^{+ \infty } _{- 4.5 }  $     &   $  0.36 ^{+ 0.06 } _{- 0.09 }  $     &   $  5.4 ^{+ 0.3 } _{- 0.2 }  $    &   $  7.8 ^{+ 0.4 } _{- 0.3 }  $   &   143.4/118 
 \\    0742590301  &   $  1.68 ^{+ 0.23 } _{- 0.20 }  $     &   $  0.179 ^{+ 0.012 } _{- 0.012 }  $     &   $  52.7 ^{+ 6.7 } _{- 5.5 }  $      &   $  2.29 ^{+ 0.11 } _{- 0.07 }  $    &   $  3.4 ^{+ 2.4 } _{- 0.7 }  $     &   $  0.35 ^{+ 0.04 } _{- 0.04 }  $     &   $  11.4 ^{+ 0.7 } _{- 0.3 }  $    &   $  16.2 ^{+ 1.0 } _{- 0.8 }  $   &   173.2/133 
 \\    0782310101  &   $  1.62 ^{+ 0.21 } _{- 0.18 }  $     &   $  0.180 ^{+ 0.012 } _{- 0.012 }  $     &   $  45.8 ^{+ 5.7 } _{- 4.6 }  $      &   $  2.33 ^{+ 0.04 } _{- 0.03 }  $    &          100     &   $  0.32 ^{+ 0.03 } _{- 0.03 }  $     &   $  10.4 ^{+ 0.4 } _{- 0.6 }  $    &   $  14.1 ^{+ 0.9 } _{- 0.2 }  $   &   165.5/135 
 \\    0794580601  &   $  1.86 ^{+ 0.47 } _{- 0.33 }  $     &   $  0.169 ^{+ 0.017 } _{- 0.012 }  $     &   $  39.8 ^{+ 12.5 } _{- 11.6 }  $      &   $  1.91 ^{+ 0.05 } _{- 0.05 }  $    &   $  2.7 ^{+ 0.6 } _{- 0.4 }  $     &   $  0.21 ^{+ 0.11 } _{- 0.12 }  $     &   $  8.2 ^{+ 1.5 } _{- 1.0 }  $    &   $  11.6 ^{+ 1.3 } _{- 0.9 }  $   &    93.2/104 
 \\    0803990201  &   $  1.59 ^{+ 0.15 } _{- 0.15 }  $     &   $  0.181 ^{+ 0.010 } _{- 0.010 }  $     &   $  43.9 ^{+ 3.7 } _{- 3.6 }  $      &   $  2.17 ^{+ 0.08 } _{- 0.04 }  $    &   $  5.1 ^{+ 33.1 } _{- 1.3 }  $     &   $  0.31 ^{+ 0.04 } _{- 0.04 }  $     &   $  9.7 ^{+ 0.5 } _{- 0.4 }  $    &   $  13.5 ^{+ 0.5 } _{- 0.4 }  $   &   191.5/149 
 \\    0803990301  &   $  1.63 ^{+ 0.22 } _{- 0.20 }  $     &   $  0.174 ^{+ 0.010 } _{- 0.012 }  $     &   $  34.8 ^{+ 4.3 } _{- 5.6 }  $      &   $  1.73 ^{+ 0.04 } _{- 0.03 }  $    &   $  2.6 ^{+ 0.3 } _{- 0.2 }  $     &   $  0.30 ^{+ 0.06 } _{- 0.15 }  $     &   $  6.3 ^{+ 0.6 } _{- 0.2 }  $    &   $  9.3 ^{+ 0.5 } _{- 0.4 }  $   &     162/128 
 \\    0803990401  &   $  1.46 ^{+ 0.15 } _{- 0.20 }  $     &   $  0.168 ^{+ 0.012 } _{- 0.007 }  $     &   $  26.6 ^{+ 4.7 } _{- 5.6 }  $      &   $  1.73 ^{+ 0.01 } _{- 0.02 }  $    &   $  2.4 ^{+ 0.2 } _{- 0.1 }  $     &   $  0.22 ^{+ 0.10 } _{- 0.11 }  $     &   $  6.5 ^{+ 0.5 } _{- 0.4 }  $    &   $  8.8 ^{+ 0.4 } _{- 0.3 }  $   &   156.6/136 
 \\    0803990501  &   $  1.57 ^{+ 0.23 } _{- 0.19 }  $     &   $  0.185 ^{+ 0.014 } _{- 0.015 }  $     &   $  40.0 ^{+ 4.8 } _{- 10.6 }  $      &   $  2.03 ^{+ 0.07 } _{- 0.06 }  $    &   $  3.4 ^{+ 2.0 } _{- 0.6 }  $     &   $  0.29 ^{+ 0.07 } _{- 0.10 }  $     &   $  8.4 ^{+ 0.8 } _{- 0.6 }  $    &   $  12.0 ^{+ 0.6 } _{- 0.5 }  $   &   123.9/129 
 \\    0803990601  &   $  1.52 ^{+ 0.15 } _{- 0.20 }  $     &   $  0.194 ^{+ 0.014 } _{- 0.010 }  $     &   $  50.8 ^{+ 4.0 } _{- 3.6 }  $      &   $  2.35 ^{+ 0.23 } _{- 0.10 }  $    &   $  4.1 ^{+ \infty } _{- 1.1 }  $     &   $  0.39 ^{+ 0.05 } _{- 0.04 }  $     &   $  10.0 ^{+ 0.6 } _{- 0.3 }  $    &   $  14.8 ^{+ 0.6 } _{- 0.5 }  $   &   179.7/136 
 \\  \cutinhead{(11) NGC 1313 X-2} 
   0405090101  &   $  1.04 ^{+ 0.24 } _{- 0.28 }  $     &   $  0.225 ^{+ 0.027 } _{- 0.021 }  $     &   $  15.2 ^{+ 1.5 } _{- 1.4 }  $      &   $  2.28 ^{+ 0.15 } _{- 0.28 }  $    &          100     &   $  0.58 ^{+ 0.04 } _{- 0.04 }  $     &   $  7.1 ^{+ 0.2 } _{- 0.2 }  $    &   $  8.5 ^{+ 0.3 } _{- 0.2 }  $   &   125.2/130 
 \\    0693850501  &   $  0.86 ^{+ 0.22 } _{- 0.20 }  $     &   $  0.235 ^{+ 0.022 } _{- 0.020 }  $     &   $  12.7 ^{+ 1.2 } _{- 1.1 }  $      &   $  3.09 ^{+ 0.27 } _{- 0.69 }  $    &          100     &   $  0.57 ^{+ 0.05 } _{- 0.08 }  $     &   $  4.6 ^{+ 0.2 } _{- 0.1 }  $    &   $  5.9 ^{+ 0.2 } _{- 0.2 }  $   &   112.4/114 
 \\    0764770101  &   $  1.03 ^{+ 0.19 } _{- 0.22 }  $     &   $  0.219 ^{+ 0.009 } _{- 0.017 }  $     &   $  6.1 ^{+ 0.6 } _{- 1.0 }  $      &   $  1.66 ^{+ 0.10 } _{- 0.08 }  $    &   $  2.1 ^{+ 0.6 } _{- 0.2 }  $     &   $  0.01 ^{+ 0.16 } _{- 0.01 }  $     &   $  1.8 ^{+ 0.1 } _{- 0.1 }  $    &   $  2.3 ^{+ 0.1 } _{- 0.2 }  $   &     86.1/81 
 \\  \cutinhead{(12) NGC 300 ULX1} 
   0791010101  &   $  0.00 ^{+ 0.02 } _{- 0.00 }  $     &   $  0.185 ^{+ 0.003 } _{- 0.003 }  $     &   $  2.9 ^{+ 0.2 } _{- 0.1 }  $      &   $  1.54 ^{+ 0.01 } _{- 0.01 }  $    &   $  2.0 ^{+ 0.0 } _{- 0.1 }  $     &   $  0.02 ^{+ 0.04 } _{- 0.02 }  $     &   $  2.17 ^{+ 0.01 } _{- 0.01 }  $    &   $  2.43 ^{+ 0.01 } _{- 0.02 }  $   &   190.7/147 
 \\    0791010301  &   $  0.00 ^{+ 0.01 } _{- 0.00 }  $     &   $  0.193 ^{+ 0.004 } _{- 0.004 }  $     &   $  3.0 ^{+ 0.2 } _{- 0.2 }  $      &   $  1.53 ^{+ 0.02 } _{- 0.02 }  $    &   $  1.9 ^{+ 0.1 } _{- 0.1 }  $     &   $  0.01 ^{+ 0.06 } _{- 0.01 }  $     &   $  2.22 ^{+ 0.02 } _{- 0.02 }  $    &   $  2.49 ^{+ 0.02 } _{- 0.02 }  $   &     107/122 
 \\ \cutinhead{(13) Holmberg IX X-1} 
   0112521001  &   $  0.68 ^{+ 0.28 } _{- 0.39 }  $     &   $  0.220 ^{+ 0.037 } _{- 0.033 }  $     &   $  14.5 ^{+ 4.3 } _{- 4.5 }  $      &   $  1.71 ^{+ 0.05 } _{- 0.07 }  $    &   $  2.7 ^{+ 0.9 } _{- 0.4 }  $     &   $  0.09 ^{+ 0.06 } _{- 0.09 }  $     &   $  13.4 ^{+ 0.8 } _{- 1.0 }  $    &   $  14.8 ^{+ 0.9 } _{- 1.1 }  $   &     71.8/88 
 \\    0200980101  &   $  0.87 ^{+ 0.05 } _{- 0.13 }  $     &   $  0.178 ^{+ 0.009 } _{- 0.006 }  $     &   $  21.2 ^{+ 2.0 } _{- 2.4 }  $      &   $  1.57 ^{+ 0.01 } _{- 0.02 }  $    &   $  2.4 ^{+ 0.1 } _{- 0.1 }  $     &   $  0.05 ^{+ 0.03 } _{- 0.05 }  $     &   $  11.4 ^{+ 0.1 } _{- 0.2 }  $    &   $  13.1 ^{+ 0.3 } _{- 0.2 }  $   &   167.3/143 
 \\    0693851101  &   $  0.63 ^{+ 0.44 } _{- 0.37 }  $     &   $  0.201 ^{+ 0.029 } _{- 0.024 }  $     &   $  55.4 ^{+ 18.4 } _{- 11.4 }  $      &   $  2.26 ^{+ 0.14 } _{- 0.11 }  $    &          100     &   $  0.56 ^{+ 0.06 } _{- 0.05 }  $     &   $  30.9 ^{+ 1.0 } _{- 0.8 }  $    &   $  36.0 ^{+ 1.0 } _{- 1.4 }  $   &      100/96 
 \\    0693851701  &   $  0.60 ^{+ 0.46 } _{- 0.35 }  $     &   $  0.209 ^{+ 0.034 } _{- 0.031 }  $     &   $  53.0 ^{+ 13.7 } _{- 8.0 }  $      &   $  2.01 ^{+ 0.42 } _{- 0.17 }  $    &   $  2.6 ^{+ \infty } _{- 0.5 }  $     &   $  0.50 ^{+ 0.11 } _{- 0.10 }  $     &   $  27.8 ^{+ 1.1 } _{- 1.1 }  $    &   $  32.3 ^{+ 0.9 } _{- 0.8 }  $   &   104.9/113 
 \\    0693851801  &   $  0.44 ^{+ 0.22 } _{- 0.19 }  $     &   $  0.222 ^{+ 0.016 } _{- 0.014 }  $     &   $  57.5 ^{+ 6.4 } _{- 5.2 }  $      &   $  2.44 ^{+ 0.27 } _{- 0.36 }  $    &   $  5.6 ^{+ \infty } _{- 3.2 }  $     &   $  0.66 ^{+ 0.05 } _{- 0.07 }  $     &   $  29.1 ^{+ 0.5 } _{- 0.5 }  $    &   $  35.3 ^{+ 0.6 } _{- 0.8 }  $   &   110.9/123 
 \\  \cutinhead{(14) NGC 2403 X-1} 
   0164560901  &   $  0.70 ^{+ 0.37 } _{- 0.32 }  $     &   $  0.249 ^{+ 0.032 } _{- 0.031 }  $     &   $  21.0 ^{+ 39.8 } _{- 4.1 }  $      &   $  4.16 ^{+ 0.68 } _{- 1.31 }  $    &          100     &   $  0.58 ^{+ 0.09 } _{- 0.07 }  $     &   $  5.7 ^{+ 1.2 } _{- 0.9 }  $    &   $  3.2 ^{+ 0.2 } _{- 0.1 }  $   &     78.7/84 
 \\    \cutinhead{(15) IC 342 X-1} 
   0206890201  &   $  2.67 ^{+ 1.21 } _{- 0.91 }  $     &   $  0.238 ^{+ 0.051 } _{- 0.045 }  $     &   $  8.0 ^{+ 9.3 } _{- 2.9 }  $      &   $  1.70 ^{+ 0.08 } _{- 0.08 }  $    &          100     &   $  0.21 ^{+ 0.25 } _{- 0.21 }  $     &   $  5.1 ^{+ 0.5 } _{- 1.0 }  $    &   $  5.4 ^{+ 1.2 } _{- 0.5 }  $   &     65.4/78 
 \\    0693850601  &   $  3.28 ^{+ 0.93 } _{- 0.97 }  $     &   $  0.191 ^{+ 0.036 } _{- 0.025 }  $     &   $  14.5 ^{+ 11.2 } _{- 7.1 }  $      &   $  1.75 ^{+ 0.05 } _{- 0.06 }  $    &   $  3.3 ^{+ 2.4 } _{- 0.6 }  $     &   $  0.23 ^{+ 0.25 } _{- 0.23 }  $     &   $  4.6 ^{+ 1.1 } _{- 0.5 }  $    &   $  6.4 ^{+ 1.3 } _{- 0.9 }  $   &     98.9/97 
 \\ 
 \enddata
\tablecomments{
Column~1: Observation ID.
Column~2: Absorption column density beyond the Milky Way; zero means that no additional absorption is needed.
Column~3: Blackbody temperature.
Column~4: Blackbody luminosity.
Column~5: Power-law Photon index.
Column~6: Electron temperature; fixed at 100 keV if the spectrum shows no downturn.
Column~7: Seed photon temperature .
Column~8: Luminosity of the Comptonization component in 0.3--10 keV.
Column~9: Total X-ray luminosity in 0.3--10 keV.
Column~10: $\chi^2$ over degrees of freedom.
 }
\end{deluxetable*}

%% file: superwind_ulx_v6_sub.bbl
\begin{thebibliography}{}
\expandafter\ifx\csname natexlab\endcsname\relax\def\natexlab#1{#1}\fi
\providecommand{\url}[1]{\href{#1}{#1}}
\providecommand{\dodoi}[1]{doi:~\href{http://doi.org/#1}{\nolinkurl{#1}}}
\providecommand{\doeprint}[1]{\href{http://ascl.net/#1}{\nolinkurl{http://ascl.net/#1}}}
\providecommand{\doarXiv}[1]{\href{https://arxiv.org/abs/#1}{\nolinkurl{https://arxiv.org/abs/#1}}}

\bibitem[{{Abolmasov} {et~al.}(2007){Abolmasov}, {Fabrika}, {Sholukhova}, \&
  {Afanasiev}}]{Abolmasov2007}
{Abolmasov}, P., {Fabrika}, S., {Sholukhova}, O., \& {Afanasiev}, V. 2007,
  Astrophysical Bulletin, 62, 36, \dodoi{10.1134/S199034130701004X}

\bibitem[{{Anand} {et~al.}(2018){Anand}, {Rizzi}, \& {Tully}}]{Anand2018}
{Anand}, G.~S., {Rizzi}, L., \& {Tully}, R.~B. 2018, \aj, 156, 105,
  \dodoi{10.3847/1538-3881/aad3b2}

\bibitem[{{Bachetti} {et~al.}(2014){Bachetti}, {Harrison}, {Walton},
  {Grefenstette}, {Chakrabarty}, {F{\"u}rst}, {Barret}, {Beloborodov}, {Boggs},
  {Christensen}, {Craig}, {Fabian}, {Hailey}, {Hornschemeier}, {Kaspi},
  {Kulkarni}, {Maccarone}, {Miller}, {Rana}, {Stern}, {Tendulkar}, {Tomsick},
  {Webb}, \& {Zhang}}]{Bachetti2014}
{Bachetti}, M., {Harrison}, F.~A., {Walton}, D.~J., {et~al.} 2014, \nat, 514,
  202, \dodoi{10.1038/nature13791}

\bibitem[{{Barnard}(2010)}]{Barnard2010}
{Barnard}, R. 2010, \mnras, 404, 42, \dodoi{10.1111/j.1365-2966.2010.16291.x}

\bibitem[{{Belfiore} {et~al.}(2020){Belfiore}, {Esposito}, {Pintore}, {Novara},
  {Salvaterra}, {De Luca}, {Tiengo}, {Caraveo}, {F{\"u}rst}, {Israel},
  {Magistrali}, {Marelli}, {Mereghetti}, {Papitto}, {Rodr{\'\i}guez Castillo},
  {Salvaggio}, {Stella}, {Walton}, {Wolter}, \& {Zampieri}}]{Belfiore2020}
{Belfiore}, A., {Esposito}, P., {Pintore}, F., {et~al.} 2020, Nature Astronomy,
  4, 147, \dodoi{10.1038/s41550-019-0903-z}

\bibitem[{{Blaes} {et~al.}(2011){Blaes}, {Krolik}, {Hirose}, \&
  {Shabaltas}}]{Blaes2011}
{Blaes}, O., {Krolik}, J.~H., {Hirose}, S., \& {Shabaltas}, N. 2011, \apj, 733,
  110, \dodoi{10.1088/0004-637X/733/2/110}

\bibitem[{{Carpano} {et~al.}(2018){Carpano}, {Haberl}, {Maitra}, \&
  {Vasilopoulos}}]{Carpano2018}
{Carpano}, S., {Haberl}, F., {Maitra}, C., \& {Vasilopoulos}, G. 2018, \mnras,
  476, L45, \dodoi{10.1093/mnrasl/sly030}

\bibitem[{{Erkut} {et~al.}(2020){Erkut}, {T{\"u}rko{\u{g}}lu}, {Ek{\c s}i}, \&
  {Alpar}}]{Erkut2020}
{Erkut}, M.~H., {T{\"u}rko{\u{g}}lu}, M.~M., {Ek{\c s}i}, K.~Y., \& {Alpar},
  M.~A. 2020, \apj, 899, 97, \dodoi{10.3847/1538-4357/aba61b}

\bibitem[{{Feng} {et~al.}(2016){Feng}, {Tao}, {Kaaret}, \&
  {Gris{\'e}}}]{Feng2016}
{Feng}, H., {Tao}, L., {Kaaret}, P., \& {Gris{\'e}}, F. 2016, \apj, 831, 117,
  \dodoi{10.3847/0004-637X/831/2/117}

\bibitem[{{F{\"u}rst} {et~al.}(2016){F{\"u}rst}, {Walton}, {Harrison}, {Stern},
  {Barret}, {Brightman}, {Fabian}, {Grefenstette}, {Madsen}, {Middleton},
  {Miller}, {Pottschmidt}, {Ptak}, {Rana}, \& {Webb}}]{Fuerst2016}
{F{\"u}rst}, F., {Walton}, D.~J., {Harrison}, F.~A., {et~al.} 2016, \apjl, 831,
  L14, \dodoi{10.3847/2041-8205/831/2/L14}

\bibitem[{{Gieren} {et~al.}(2005){Gieren}, {Pietrzy{\'n}ski}, {Soszy{\'n}ski},
  {Bresolin}, {Kudritzki}, {Minniti}, \& {Storm}}]{Gieren2005}
{Gieren}, W., {Pietrzy{\'n}ski}, G., {Soszy{\'n}ski}, I., {et~al.} 2005, \apj,
  628, 695, \dodoi{10.1086/430903}

\bibitem[{{Ho} {et~al.}(1997){Ho}, {Filippenko}, \& {Sargent}}]{Ho1997}
{Ho}, L.~C., {Filippenko}, A.~V., \& {Sargent}, W. L.~W. 1997, \apjs, 112, 315,
  \dodoi{10.1086/313041}

\bibitem[{{Israel} {et~al.}(2017{\natexlab{a}}){Israel}, {Belfiore}, {Stella},
  {Esposito}, {Casella}, {De Luca}, {Marelli}, {Papitto}, {Perri}, {Puccetti},
  {Castillo}, {Salvetti}, {Tiengo}, {Zampieri}, {D'Agostino}, {Greiner},
  {Haberl}, {Novara}, {Salvaterra}, {Turolla}, {Watson}, {Wilms}, \&
  {Wolter}}]{Israel2017}
{Israel}, G.~L., {Belfiore}, A., {Stella}, L., {et~al.} 2017{\natexlab{a}},
  Science, 355, 817, \dodoi{10.1126/science.aai8635}

\bibitem[{{Israel} {et~al.}(2017{\natexlab{b}}){Israel}, {Papitto}, {Esposito},
  {Stella}, {Zampieri}, {Belfiore}, {Rodr{\'\i}guez Castillo}, {De Luca},
  {Tiengo}, {Haberl}, {Greiner}, {Salvaterra}, {Sandrelli}, \&
  {Lisini}}]{Israel2017a}
{Israel}, G.~L., {Papitto}, A., {Esposito}, P., {et~al.} 2017{\natexlab{b}},
  \mnras, 466, L48, \dodoi{10.1093/mnrasl/slw218}

\bibitem[{{Jiang} {et~al.}(2014){Jiang}, {Stone}, \& {Davis}}]{Jiang2014}
{Jiang}, Y.-F., {Stone}, J.~M., \& {Davis}, S.~W. 2014, \apj, 796, 106,
  \dodoi{10.1088/0004-637X/796/2/106}

\bibitem[{{Kaaret} {et~al.}(2017){Kaaret}, {Feng}, \& {Roberts}}]{Kaaret2017}
{Kaaret}, P., {Feng}, H., \& {Roberts}, T.~P. 2017, \araa, 55, 303,
  \dodoi{10.1146/annurev-astro-091916-055259}

\bibitem[{{Kalberla} {et~al.}(2005){Kalberla}, {Burton}, {Hartmann}, {Arnal},
  {Bajaja}, {Morras}, \& {P{\"o}ppel}}]{Kalberla2005}
{Kalberla}, P.~M.~W., {Burton}, W.~B., {Hartmann}, D., {et~al.} 2005, \aap,
  440, 775, \dodoi{10.1051/0004-6361:20041864}

\bibitem[{{Kara} {et~al.}(2020){Kara}, {Pinto}, {Walton}, {Alston}, {Bachetti},
  {Barret}, {Brightman}, {Canizares}, {Earnshaw}, {Fabian}, {F{\"u}rst},
  {Kosec}, {Middleton}, {Roberts}, {Soria}, {Tao}, \& {Webb}}]{Kara2020}
{Kara}, E., {Pinto}, C., {Walton}, D.~J., {et~al.} 2020, \mnras, 491, 5172,
  \dodoi{10.1093/mnras/stz3318}

\bibitem[{{Karachentsev} {et~al.}(2002){Karachentsev}, {Sharina}, {Dolphin},
  {Grebel}, {Geisler}, {Guhathakurta}, {Hodge}, {Karachentseva}, {Sarajedini},
  \& {Seitzer}}]{Karachentsev2002a}
{Karachentsev}, I.~D., {Sharina}, M.~E., {Dolphin}, A.~E., {et~al.} 2002, \aap,
  385, 21, \dodoi{10.1051/0004-6361:20020042}

\bibitem[{{Karachentsev} {et~al.}(2003){Karachentsev}, {Grebel}, {Sharina},
  {Dolphin}, {Geisler}, {Guhathakurta}, {Hodge}, {Karachentseva}, {Sarajedini},
  \& {Seitzer}}]{Karachentsev2003}
{Karachentsev}, I.~D., {Grebel}, E.~K., {Sharina}, M.~E., {et~al.} 2003, \aap,
  404, 93, \dodoi{10.1051/0004-6361:20030170}

\bibitem[{{Kawashima} {et~al.}(2012){Kawashima}, {Ohsuga}, {Mineshige},
  {Yoshida}, {Heinzeller}, \& {Matsumoto}}]{Kawashima2012}
{Kawashima}, T., {Ohsuga}, K., {Mineshige}, S., {et~al.} 2012, \apj, 752, 18,
  \dodoi{10.1088/0004-637X/752/1/18}

\bibitem[{{King} \& {Pounds}(2003)}]{King2003}
{King}, A.~R., \& {Pounds}, K.~A. 2003, \mnras, 345, 657,
  \dodoi{10.1046/j.1365-8711.2003.06980.x}

\bibitem[{{Kitaki} {et~al.}(2018){Kitaki}, {Mineshige}, {Ohsuga}, \&
  {Kawashima}}]{Kitaki2018}
{Kitaki}, T., {Mineshige}, S., {Ohsuga}, K., \& {Kawashima}, T. 2018, \pasj,
  70, 108, \dodoi{10.1093/pasj/psy110}

\bibitem[{{Kosec} {et~al.}(2018){Kosec}, {Pinto}, {Walton}, {Fabian},
  {Bachetti}, {Brightman}, {F{\"u}rst}, \& {Grefenstette}}]{Kosec2018}
{Kosec}, P., {Pinto}, C., {Walton}, D.~J., {et~al.} 2018, \mnras, 479, 3978,
  \dodoi{10.1093/mnras/sty1626}

\bibitem[{{Lipunova}(1999)}]{Lipunova1999}
{Lipunova}, G.~V. 1999, Astronomy Letters, 25, 508.
\newblock \doarXiv{astro-ph/9906324}

\bibitem[{{McQuinn} {et~al.}(2016){McQuinn}, {Skillman}, {Dolphin}, {Berg}, \&
  {Kennicutt}}]{McQuinn2016}
{McQuinn}, K. B.~W., {Skillman}, E.~D., {Dolphin}, A.~E., {Berg}, D., \&
  {Kennicutt}, R. 2016, \apj, 826, 21, \dodoi{10.3847/0004-637X/826/1/21}

\bibitem[{{Meier}(1982{\natexlab{a}})}]{Meier1982_PaperII}
{Meier}, D.~L. 1982{\natexlab{a}}, \apj, 256, 681, \dodoi{10.1086/159942}

\bibitem[{{Meier}(1982{\natexlab{b}})}]{Meier1982_PaperIII}
---. 1982{\natexlab{b}}, \apj, 256, 693, \dodoi{10.1086/159943}

\bibitem[{{Meier}(2012)}]{Meier2012}
---. 2012, {Black Hole Astrophysics: The Engine Paradigm} (Berlin: Springer)

\bibitem[{{Middleton} {et~al.}(2015){Middleton}, {Heil}, {Pintore}, {Walton},
  \& {Roberts}}]{Middleton2015}
{Middleton}, M.~J., {Heil}, L., {Pintore}, F., {Walton}, D.~J., \& {Roberts},
  T.~P. 2015, \mnras, 447, 3243, \dodoi{10.1093/mnras/stu2644}

\bibitem[{{Middleton} {et~al.}(2014){Middleton}, {Walton}, {Roberts}, \&
  {Heil}}]{Middleton2014}
{Middleton}, M.~J., {Walton}, D.~J., {Roberts}, T.~P., \& {Heil}, L. 2014,
  \mnras, 438, L51, \dodoi{10.1093/mnrasl/slt157}

\bibitem[{{Miller} {et~al.}(2013){Miller}, {Walton}, {King}, {Reynolds},
  {Fabian}, {Miller}, \& {Reis}}]{Miller2013}
{Miller}, J.~M., {Walton}, D.~J., {King}, A.~L., {et~al.} 2013, \apjl, 776,
  L36, \dodoi{10.1088/2041-8205/776/2/L36}

\bibitem[{{Mushtukov} {et~al.}(2016){Mushtukov}, {Nagirner}, \&
  {Poutanen}}]{Mushtukov2016}
{Mushtukov}, A.~A., {Nagirner}, D.~I., \& {Poutanen}, J. 2016, \prd, 93,
  105003, \dodoi{10.1103/PhysRevD.93.105003}

\bibitem[{{Mushtukov} {et~al.}(2015){Mushtukov}, {Suleimanov}, {Tsygankov}, \&
  {Poutanen}}]{Mushtukov2015}
{Mushtukov}, A.~A., {Suleimanov}, V.~F., {Tsygankov}, S.~S., \& {Poutanen}, J.
  2015, \mnras, 454, 2539, \dodoi{10.1093/mnras/stv2087}

\bibitem[{{Narayan} {et~al.}(2017){Narayan}, {S{\k a}dowski}, \&
  {Soria}}]{Narayan2017}
{Narayan}, R., {S{\k a}dowski}, A., \& {Soria}, R. 2017, \mnras, 469, 2997,
  \dodoi{10.1093/mnras/stx1027}

\bibitem[{{Ohsuga} \& {Mineshige}(2011)}]{Ohsuga2011}
{Ohsuga}, K., \& {Mineshige}, S. 2011, \apj, 736, 2,
  \dodoi{10.1088/0004-637X/736/1/2}

\bibitem[{{Pakull} \& {Mirioni}(2003)}]{Pakull2003}
{Pakull}, M.~W., \& {Mirioni}, L. 2003, in Revista Mexicana de Astronomia y
  Astrofisica Conference Series, Vol.~15, Revista Mexicana de Astronomia y
  Astrofisica Conference Series, ed. J.~{Arthur} \& W.~J. {Henney}, 197--199

\bibitem[{{Pinto} {et~al.}(2016){Pinto}, {Middleton}, \& {Fabian}}]{Pinto2016}
{Pinto}, C., {Middleton}, M.~J., \& {Fabian}, A.~C. 2016, \nat, 533, 64,
  \dodoi{10.1038/nature17417}

\bibitem[{{Pinto} {et~al.}(2017){Pinto}, {Alston}, {Soria}, {Middleton},
  {Walton}, {Sutton}, {Fabian}, {Earnshaw}, {Urquhart}, {Kara}, \&
  {Roberts}}]{Pinto2017}
{Pinto}, C., {Alston}, W., {Soria}, R., {et~al.} 2017, \mnras, 468, 2865,
  \dodoi{10.1093/mnras/stx641}

\bibitem[{{Pinto} {et~al.}(2020){Pinto}, {Walton}, {Kara}, {Parker}, {Soria},
  {Kosec}, {Middleton}, {Alston}, {Fabian}, {Guainazzi}, {Roberts}, {Fuerst},
  {Earnshaw}, {Sathyaprakash}, \& {Barret}}]{Pinto2020}
{Pinto}, C., {Walton}, D.~J., {Kara}, E., {et~al.} 2020, \mnras, 492, 4646,
  \dodoi{10.1093/mnras/staa118}

\bibitem[{{Poutanen} {et~al.}(2007){Poutanen}, {Lipunova}, {Fabrika},
  {Butkevich}, \& {Abolmasov}}]{Poutanen2007}
{Poutanen}, J., {Lipunova}, G., {Fabrika}, S., {Butkevich}, A.~G., \&
  {Abolmasov}, P. 2007, \mnras, 377, 1187,
  \dodoi{10.1111/j.1365-2966.2007.11668.x}

\bibitem[{{Shakura} \& {Sunyaev}(1973)}]{Shakura1973}
{Shakura}, N.~I., \& {Sunyaev}, R.~A. 1973, \aap, 500, 33

\bibitem[{{Shen} {et~al.}(2016){Shen}, {Nakar}, \& {Piran}}]{Shen2016}
{Shen}, R.-F., {Nakar}, E., \& {Piran}, T. 2016, \mnras, 459, 171,
  \dodoi{10.1093/mnras/stw645}

\bibitem[{{S{\k{a}}dowski} \& {Narayan}(2016)}]{Sadowski2016}
{S{\k{a}}dowski}, A., \& {Narayan}, R. 2016, \mnras, 456, 3929,
  \dodoi{10.1093/mnras/stv2941}

\bibitem[{{Socrates} \& {Davis}(2006)}]{Socrates2006}
{Socrates}, A., \& {Davis}, S.~W. 2006, \apj, 651, 1049, \dodoi{10.1086/507119}

\bibitem[{{Soria} \& {Kong}(2016)}]{Soria2016}
{Soria}, R., \& {Kong}, A. 2016, \mnras, 456, 1837,
  \dodoi{10.1093/mnras/stv2671}

\bibitem[{{Tao} {et~al.}(2019){Tao}, {Feng}, {Zhang}, {Bu}, {Zhang}, {Qu}, \&
  {Zhang}}]{Tao2019}
{Tao}, L., {Feng}, H., {Zhang}, S., {et~al.} 2019, \apj, 873, 19,
  \dodoi{10.3847/1538-4357/ab0211}

\bibitem[{{Tsygankov} {et~al.}(2017){Tsygankov}, {Doroshenko}, {Lutovinov},
  {Mushtukov}, \& {Poutanen}}]{Tsygankov2017}
{Tsygankov}, S.~S., {Doroshenko}, V., {Lutovinov}, A.~A., {Mushtukov}, A.~A.,
  \& {Poutanen}, J. 2017, \aap, 605, A39, \dodoi{10.1051/0004-6361/201730553}

\bibitem[{{Tully} {et~al.}(2016){Tully}, {Courtois}, \& {Sorce}}]{Tully2016}
{Tully}, R.~B., {Courtois}, H.~M., \& {Sorce}, J.~G. 2016, \aj, 152, 50,
  \dodoi{10.3847/0004-6256/152/2/50}

\bibitem[{{Urquhart} \& {Soria}(2016)}]{Urquhart2016}
{Urquhart}, R., \& {Soria}, R. 2016, \mnras, 456, 1859,
  \dodoi{10.1093/mnras/stv2293}

\bibitem[{{Vaughan} {et~al.}(2003){Vaughan}, {Edelson}, {Warwick}, \&
  {Uttley}}]{Vaughan2003}
{Vaughan}, S., {Edelson}, R., {Warwick}, R.~S., \& {Uttley}, P. 2003, \mnras,
  345, 1271, \dodoi{10.1046/j.1365-2966.2003.07042.x}

\bibitem[{{Walton} {et~al.}(2011){Walton}, {Gladstone}, {Roberts}, {Fabian},
  {Caballero-Garcia}, {Done}, \& {Middleton}}]{Walton2011}
{Walton}, D.~J., {Gladstone}, J.~C., {Roberts}, T.~P., {et~al.} 2011, \mnras,
  414, 1011, \dodoi{10.1111/j.1365-2966.2011.18397.x}

\bibitem[{{Walton} {et~al.}(2016){Walton}, {Middleton}, {Pinto}, {Fabian},
  {Bachetti}, {Barret}, {Brightman}, {Fuerst}, {Harrison}, {Miller}, \&
  {Stern}}]{Walton2016}
{Walton}, D.~J., {Middleton}, M.~J., {Pinto}, C., {et~al.} 2016, \apjl, 826,
  L26, \dodoi{10.3847/2041-8205/826/2/L26}

\bibitem[{{Walton} {et~al.}(2020){Walton}, {Pinto}, {Nowak}, {Bachetti},
  {Sathyaprakash}, {Kara}, {Roberts}, {Soria}, {Brightman}, {Canizares},
  {Earnshaw}, {F{\"u}rst}, {Heida}, {Middleton}, {Stern}, {Tao}, {Webb},
  {Alston}, {Barret}, {Fabian}, {Harrison}, \& {Kosec}}]{Walton2020}
{Walton}, D.~J., {Pinto}, C., {Nowak}, M., {et~al.} 2020, \mnras, 494, 6012,
  \dodoi{10.1093/mnras/staa1129}

\bibitem[{{Webb} {et~al.}(2020){Webb}, {Coriat}, {Traulsen}, {Ballet}, {Motch},
  {Carrera}, {Koliopanos}, {Authier}, {de la Calle}, {Ceballos}, {Colomo},
  {Chuard}, {Freyberg}, {Garcia}, {Kolehmainen}, {Lamer}, {Lin}, {Maggi},
  {Michel}, {Page}, {Page}, {Perea-Calderon}, {Pineau}, {Rodriguez}, {Rosen},
  {Santos Lleo}, {Saxton}, {Schwope}, {Tom{\'a}s}, {Watson}, \&
  {Zakardjian}}]{Webb2020}
{Webb}, N.~A., {Coriat}, M., {Traulsen}, I., {et~al.} 2020, \aap, 641, A136,
  \dodoi{10.1051/0004-6361/201937353}

\bibitem[{{Weng} \& {Feng}(2018)}]{Weng2018}
{Weng}, S.-S., \& {Feng}, H. 2018, \apj, 853, 115,
  \dodoi{10.3847/1538-4357/aaa45c}

\bibitem[{{Yao} \& {Feng}(2019)}]{Yao2019}
{Yao}, Y., \& {Feng}, H. 2019, \apjl, 884, L3, \dodoi{10.3847/2041-8213/ab44c7}

\bibitem[{{Zdziarski} {et~al.}(1996){Zdziarski}, {Johnson}, \&
  {Magdziarz}}]{Zdziarski1996}
{Zdziarski}, A.~A., {Johnson}, W.~N., \& {Magdziarz}, P. 1996, \mnras, 283,
  193, \dodoi{10.1093/mnras/283.1.193}

\bibitem[{{Zhou} {et~al.}(2019){Zhou}, {Feng}, {Ho}, \& {Yao}}]{Zhou2019}
{Zhou}, Y., {Feng}, H., {Ho}, L.~C., \& {Yao}, Y. 2019, \apj, 871, 115,
  \dodoi{10.3847/1538-4357/aaf724}

\bibitem[{{{\.Z}ycki} {et~al.}(1999){{\.Z}ycki}, {Done}, \&
  {Smith}}]{Zycki1999}
{{\.Z}ycki}, P.~T., {Done}, C., \& {Smith}, D.~A. 1999, \mnras, 309, 561,
  \dodoi{10.1046/j.1365-8711.1999.02885.x}

\end{thebibliography}
